\documentstyle[twocolumn,floats,aps,epsf]{revtex}

\def\ltsim{\vbox {\hbox{\lower 0.9\baselineskip \hbox{$<$}} \break
         \hbox{\lower 0.2\baselineskip \hbox{$\sim$}} } }

\def\gtsim{\vbox {\hbox{\lower 0.9\baselineskip \hbox{$>$}} \break
         \hbox{\lower 0.2\baselineskip \hbox{$\sim$}} } }

\begin{document}
\def\no{\noindent}
\twocolumn[\hsize\textwidth\columnwidth\hsize\csname@twocolumnfalse%
\endcsname
\title{Density of states ``width parity" effect in  $d$-wave 
superconducting quantum wires
}
\author{K. Ziegler$^1$,
W.A. Atkinson$^{2,3}$, P.J. Hirschfeld$^2$\\
$^1$Institut f\"ur Physik, Universit\"at Augsburg, 86135 Augsburg, Germany\\
$^2$Department of Physics, University of Florida, PO Box 118440, 
Gainesville FL 32611 USA\\
$^3$Department of Physics,  Southern llinois University, 
Carbondale, IL 62901 USA.  }
\date{\today}
\maketitle
\begin{abstract}
We calculate the density of states (DOS) in a clean mesoscopic
$d$-wave superconducting quantum wire, i.e. a sample of infinite
length but finite width $N$.  For open boundary conditions and
half-filling, the DOS at zero energy is found to be zero if $N$ is
even, and nonzero if $N$ is odd.  At finite chemical potential, all
chains are gapped but the qualtitative differences between even and
odd $N$ remain.
\end{abstract}

]
\narrowtext
\section{Introduction}
In recent years the fabrication of nanoscale devices has heightened
interest in solid-state quantum systems where the discreteness of the
level spacing plays an important role.  Originally discussed only in
semiconductor quantum dots and quantum wires, following the work of
Ralph et al \cite{Ralph1} many fascinating consequences of the level
discreteness were also observed in ultrasmall metal particles.  These
included the observation of a spectroscopic gap attributed to pair
correlations in the small grains which depended on the so-called
"number parity" of the sample, i.e. whether the number of electrons in
the grain was even or odd.  The physics of the levels in such grains
has been quite well understood.\cite{vonDelft} Superconducting systems
extended in one dimension but mesoscopic in another have also been
studied.\cite{Zaikin}\vskip .2cm 
In most of these studies, a gap with
$A_{1g}$ or ``s-wave" symmetry has been assumed for simplicity, unless
otherwise stated.  However, for some practical purposes it may be of
interest to use cuprate samples, which are known to have $d_{x^2-y^2}$
symmetry.  From a fundamental point of view it is also expected that
these systems will be different because already in the bulk state
there are states below the maximum gap scale; the pure $d$-wave bulk
DOS is linear in energy, $\rho(E)\sim |E|$.  We have studied the DOS
of mesoscopic $d$-wave quantum wires and found a new kind of "parity
effect", not related to the number of particles in the system, but to
the parity of the number of ``chains" across the mesoscopic sample.
In some ways, the effect is reminiscent of a type of mesoscopic gap
effect which has been recently discussed in the context of single-wall
carbon nanotubes, where the existence or nonexistence of a gap depends
on the intrinsic twist or chirality induced while forming the
tube.\cite{nanotube} The $d$-wave systems we discuss, while they may
be considered to be ``tubes" if periodic boundary conditions are
employed, do not break either time reversal symmetry or spatial
parity.  Nevertheless, at half-filling,  
we observe that if the width of the system $N$
is even there is a gap in the DOS, whereas if the width is odd the
density of states at zero energy $\rho(0)$ is nonzero.  As $N$ is
increased, both the even-$N$ gap and the odd-$N$ $\rho(0)$ vanish as
they must to give the bulk $d$-wave result $\rho(E)\sim |E|$.  
Although the DOS vanishes at $E=0$ for all $N$ when $\mu \neq 0$,
there is still a pronounced even-odd effect in the DOS, and we
argue that this effect should be observable in quantum wires
fabricated from cuprate
superconductors.   We speculate further that it may
be of relevance to the study of disorder-induced pseudogaps in bulk
$d$-wave superconductors, which have been the subject of much
controversy in recent years.\cite{2Ddos}

\section{Model}
We consider the DOS of a $d$-wave superconducting chain (DWSC) that
is coupled to $N-1$ other DWSCs. The chains form a 2D system, and the
first chain is at the boundary. Similar problems were considered
in previous studies of random flux systems \cite{miller} and
quasi-one-dimensional disordered tight-binding models
\cite{brouwer1,brouwer2}. In this paper we will ignore the effect
of disorder and study the odd-even effect in a pure system of
coupled DWSCs. Such a behavior is also known for a system of coupled
spin chains which has a gap (no gap) if the number of chains is even (odd)
\cite{haldane,fradkin}.

There are several possible geometries one can assume for this problem.
The most natural, and the one we consider here, is a wire parallel to
the 100 (or 010) crystal direction.  In this geometry, the boundaries
of the wire are aligned with both the underlying CuO$_2$ crystal and
the gap maxima (which are tied to the crystal lattice).  Other
geometries, such as a 110 oriented wire, are expected to show similar
even-odd effects to the ones described below, but these will be
mixed with Andreev scattering resonances\cite{sauls} and thus
will be more difficult to interpret.  In addition, we anticipate that
the 100 oriented wire is the most accessible to current technology.

The Bogoliubov-de Gennes Hamiltonian of the 2D DWS
\[
H=(-\nabla^2+\mu)\sigma_3+\Delta\sigma_1
\]
is defined on a square lattice with
$$
-\nabla^2f(r)=-t\sum_{j=1}^2[f(r+e_j)+f(r-e_j)]
$$
and the DWS order parameter
$$
\Delta f(r)=\Delta_0\sum_{j=1}^2(-1)^j[f(r+e_j)+f(r-e_j)].
$$
With this definition, the gap in familiar $k$-space representation
is $\Delta_k = 2\Delta_0[\cos(k_x) - \cos(k_y)]$.
In the following calculations we will measure energies in units of $t$
for simplicity.  Moreover, it is assumed that chains are infinitely
long and arranged parallel to the $y$-axis of our 2D system.

\section{Periodic Boundaries}

The finite set of chains can be closed periodically in the
$x$-direction by identifying the first with the $N+1^{\rm th}$ chain.
While this is not particularly physical, the calculation is
fairly transparent and therefore worthwile for pedagogical reasons.
For this translational-invariant $N$-chain system the
integration in $y$-direction (i.e. along the chains) can be performed
and gives the DOS at zero energy
\begin{eqnarray}
\rho(E=0)&=& -{1\over\pi N} \sum_{k_x}\int_{-\pi}^{\pi} {dk_y\over
2\pi}
\,{\rm Im}\,G_0({\bf k},+i\epsilon) 
\nonumber\\
&=&\lim_{\epsilon\to0}{\epsilon\rho_0\over N}\sum_{n=0}^{N-1}
{\displaystyle 1\over\sqrt{\displaystyle \epsilon^2+ [\mu+4t\cos(k_x)]^2
}},
\label{pdos}
\end{eqnarray}
where 
\[
\rho_0={2t\over \pi\sqrt{t^2+\Delta_0^2}\sqrt{16t^2-\mu^2}}.
\]
Here $G_0$ is the Green's function for a clean 2D $d$-wave
superconductor,  $k_x=2\pi n/N$ ($n=0,1,... ,N-1$), and we have
set the lattice constant to unity.  First, we consider
the DOS at half-filling ($\mu=0$).  We find that 
$\rho(E=0)$ is {\em finite} for $N$ a multiple of 4 since
the allowed values of $k_x$ include
$k_x = \pi/2$ and $3\pi/2$ which produce finite contributions to
Eq.~(\ref{pdos}).
For other values of $N$, on the other hand,
expression (\ref{pdos}) always vanishes.
At $\mu=0$, then,  
we have
$$
\rho_N(E=0)=\cases{
2\rho_0/ N & for $N=4k$, $k=1,2,...$ \cr
0 & otherwise
}.
$$
This periodicity in $N$ reflects the conditions necessary to form a
standing wave in the $x$-direction at the Fermi energy. For $\mu=0$,
it is the four nodal points at ${\bf k} = (\pm \pi/2, \pm \pi/2)$
which produce the standing wave.  The wave function at $E=0$
is thus exactly the same as in the normal state;
it is therefore not surprising that 
$\rho(E=0)$ is only weakly dependent on $\Delta_0$.

For given N,  there will be a discrete set of $\mu$ which
produce a finite density of states at $E=0$.  For generic 
$\mu$, the nodal eigenstates do not contribute
at  $E=0$, but we expect on the basis of (1) that 
a qualitative width parity effect will remain at nonzero $E$.
As we shall see in the next section, numerical studies on
 realistic models find a pronounced effect of this type.

\section{Open Boundaries}

Periodic boundary conditions are not realistic for typical experiments
because the superconductor is a planar object. Therefore, a better
choice is open boundaries if we have a situation with a finite number
of chains. The DOS is calculated numerically for
$N=4,5,20,21$, as shown in Figs.  1-5.  Since we have just
shown that $\Delta_0$ plays a minor role in the parity effect, we
proceed (for simplicity) under the assumption $\Delta_0/t=1$ unless
explicitly stated otherwise.  In numerical work (e.g. Fig.\ 6 below), it is
shown that this assumption does not change the qualitative results.
From the figures the alternating structure of the DOS for odd and even
number of chains is immediately obvious, and may be characterized by
the DOS of the end chain.  


\subsection{Recursive method}
Some insight into the dependence of these gaps and residual DOS's 
on system size and other paramters may be
obtained from analytical methods. 
First, the Hamiltonian $H$ can be diagonalized with respect to $y$-direction
by a Fourier transformation $y\to k_y$. Then the structure with respect
to the $x$-direction can be expressed in a matrix form for $x=1,2,...,N$
as
$$
{\hat H}_N=\pmatrix{
H_{NN} & H_{NN-1} & 0 & \ldots & 0 \cr
H_{N-1N} & H_{N-1N-1} & \ddots & \ddots & \vdots \cr
0      & \ddots & \ddots & \ddots & 0 \cr
\vdots & \ddots & \ddots & H_{22} & H_{21} \cr
0      & \ldots & 0 & H_{12} & H_{11} \cr
}.
$$
For a compact notation we introduce a block matrix representation
$$
{\hat H}_N:=\pmatrix{
H_{NN} & {\hat H}_{NN-1} \cr
{\hat H}_{N-1N} & {\hat H}_{N-1} \cr
}
$$
with
\begin{equation}
{\hat H}_{NN-1}=(H_{NN-1}, 0, \ldots, 0),
\label{matrix1}
\end{equation}
\begin{equation}
{\hat H}_{N-1N}=\pmatrix{
H_{N-1N}\cr
0 \cr
\vdots \cr
0 \cr
}.
\label{matrix2}
\end{equation}

\begin{figure}[h]
\begin{picture}(100,200)
\leavevmode\centering\includegraphics{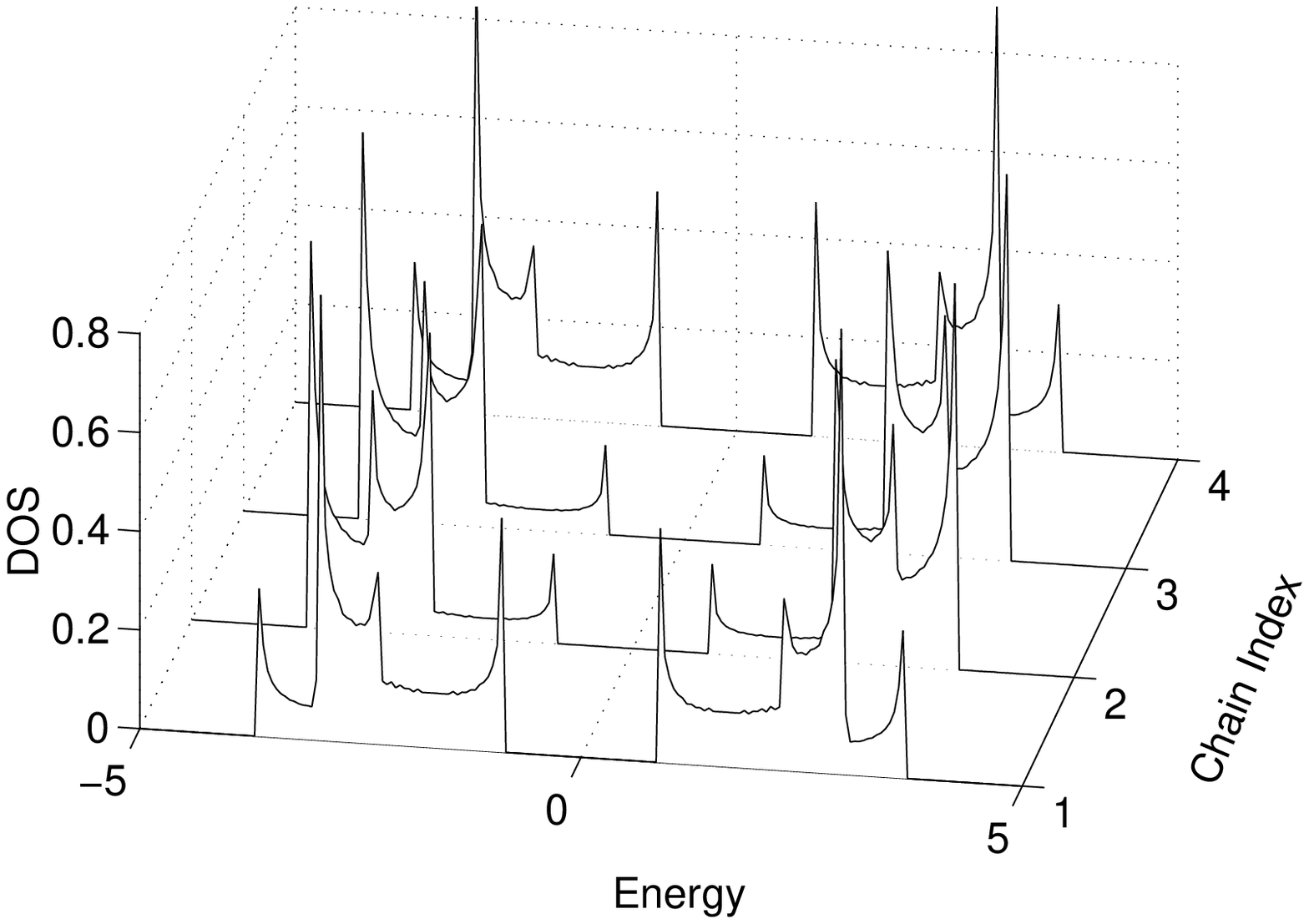}
              \end{picture}
\vskip -1cm 
\begin{picture}(100,200)
\leavevmode\centering\includegraphics{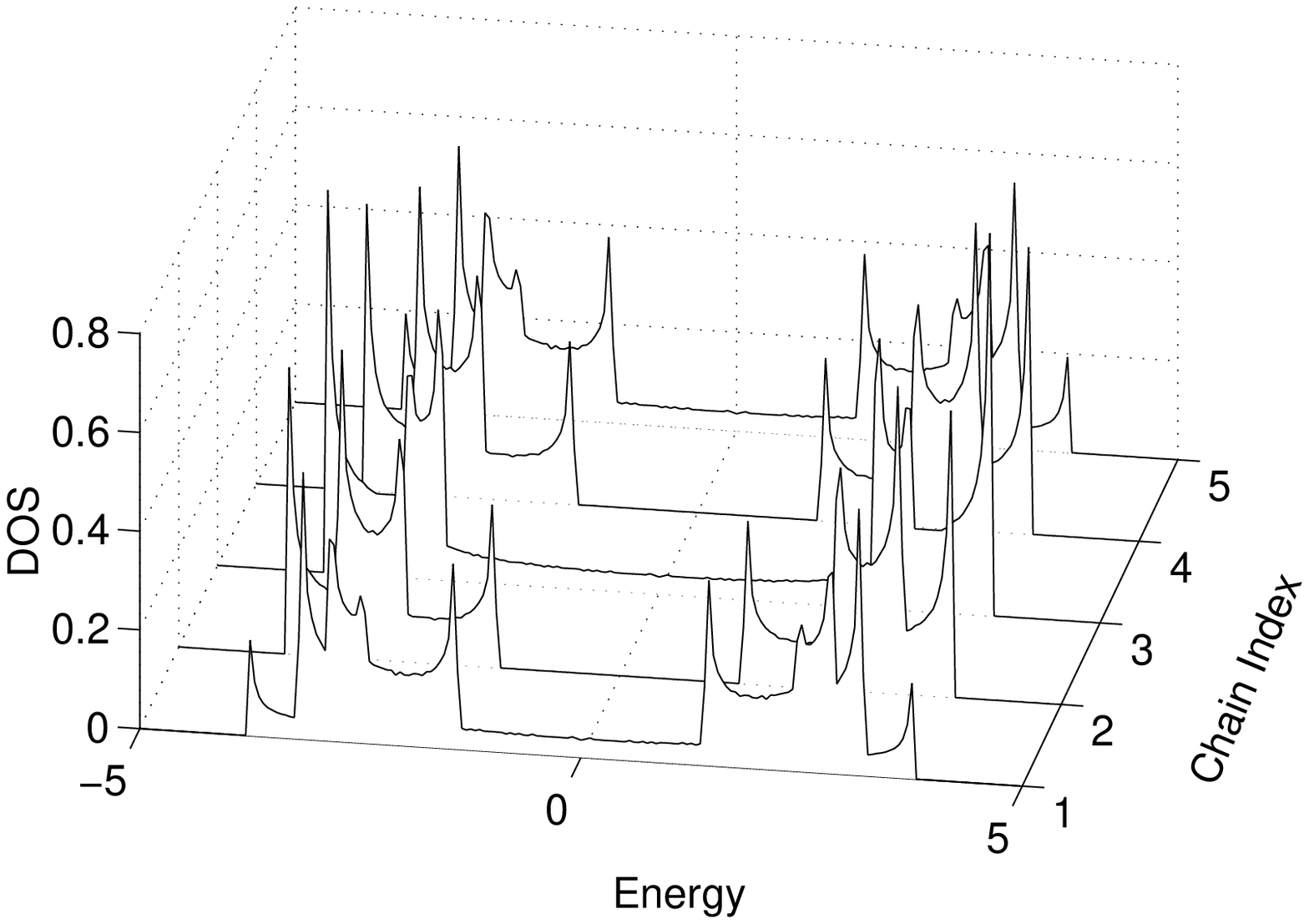}
              \end{picture}
\caption{
Density of states for four and five  chains with $\Delta_0/t=1$ and
$\mu=0$.}
\end{figure}
The corresponding Green's function at zero energy, 
which is required for the evaluation
of the DOS, can then be written (after a shift $H\to H+i\epsilon\sigma_0$)
as
$$
[{\hat H}^{-1}_N]_{NN}=(H_{NN} - {\hat H}_{NN-1}{\hat H}^{-1}_{N-1}
{\hat H}_{N-1N})^{-1}
$$
$$
=(H_{NN} - H_{NN-1}[{\hat H}^{-1}_{N-1}]_{N-1N-1}H_{N-1N})^{-1}.
$$

The second equation is due to the special form of the matrices given in
(\ref{matrix1}) and (\ref{matrix2}).
Using $\Gamma_N=[{\hat H}_N^{-1}]_{NN}$ the recurrence relation reads
\begin{equation}
\Gamma_N=(H_{NN} - H_{NN-1}\Gamma_{N-1}H_{N-1N})^{-1}.
\label{recursion}
\end{equation}
This is a ``continued fraction representation'' of
the $2\times2$ matrix $\Gamma_N$, the Green's function of end chain labeled
by $N$ at a given $k_y$. It can be used even in the presence of random
terms along the $x$-direction because no diagonalization of the matrix
is necessary.
\begin{figure}[h]
\begin{picture}(100,200)
\leavevmode\centering\includegraphics{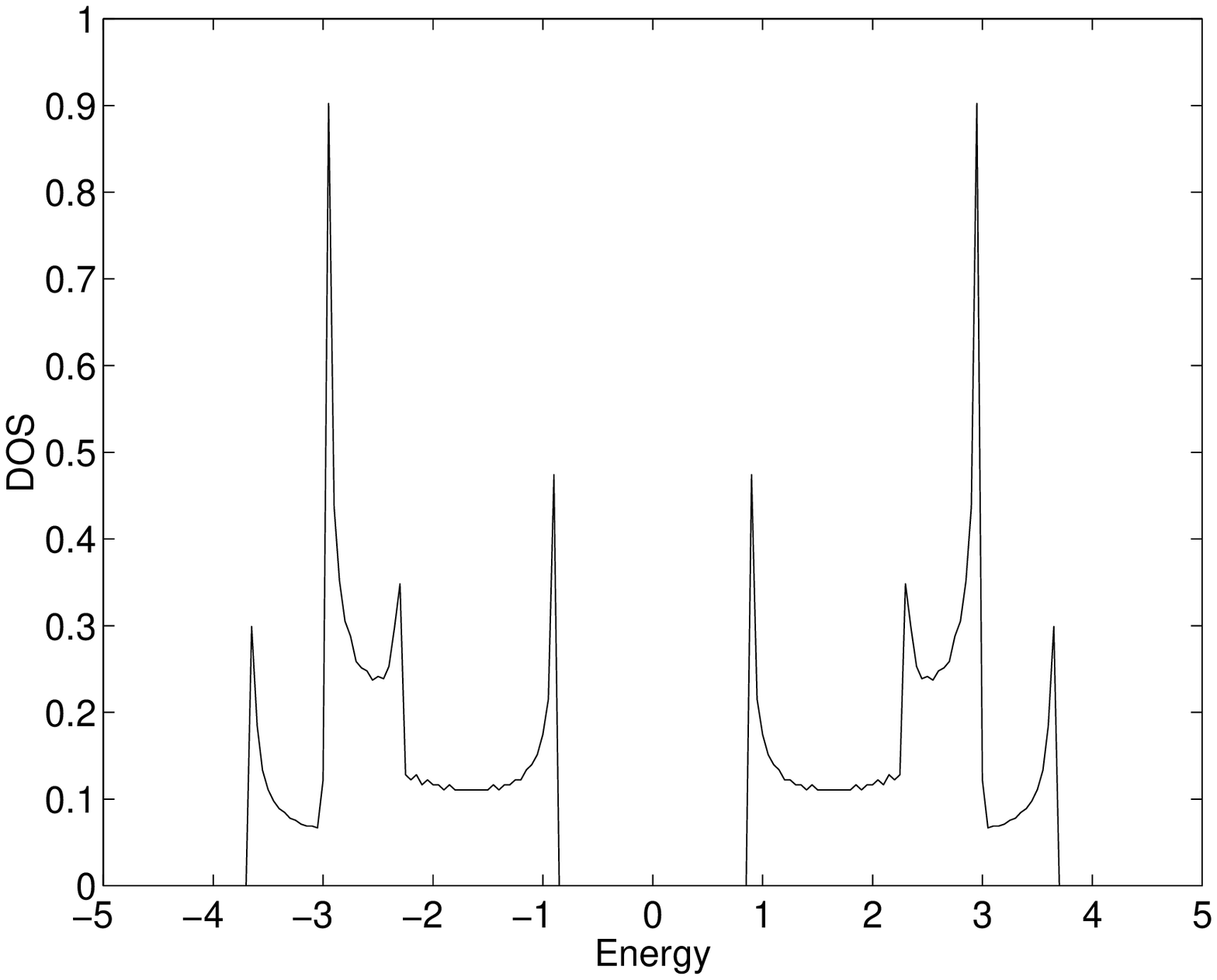}
              \end{picture}
\vskip -1cm
\begin{picture}(100,200)
\leavevmode\centering\includegraphics{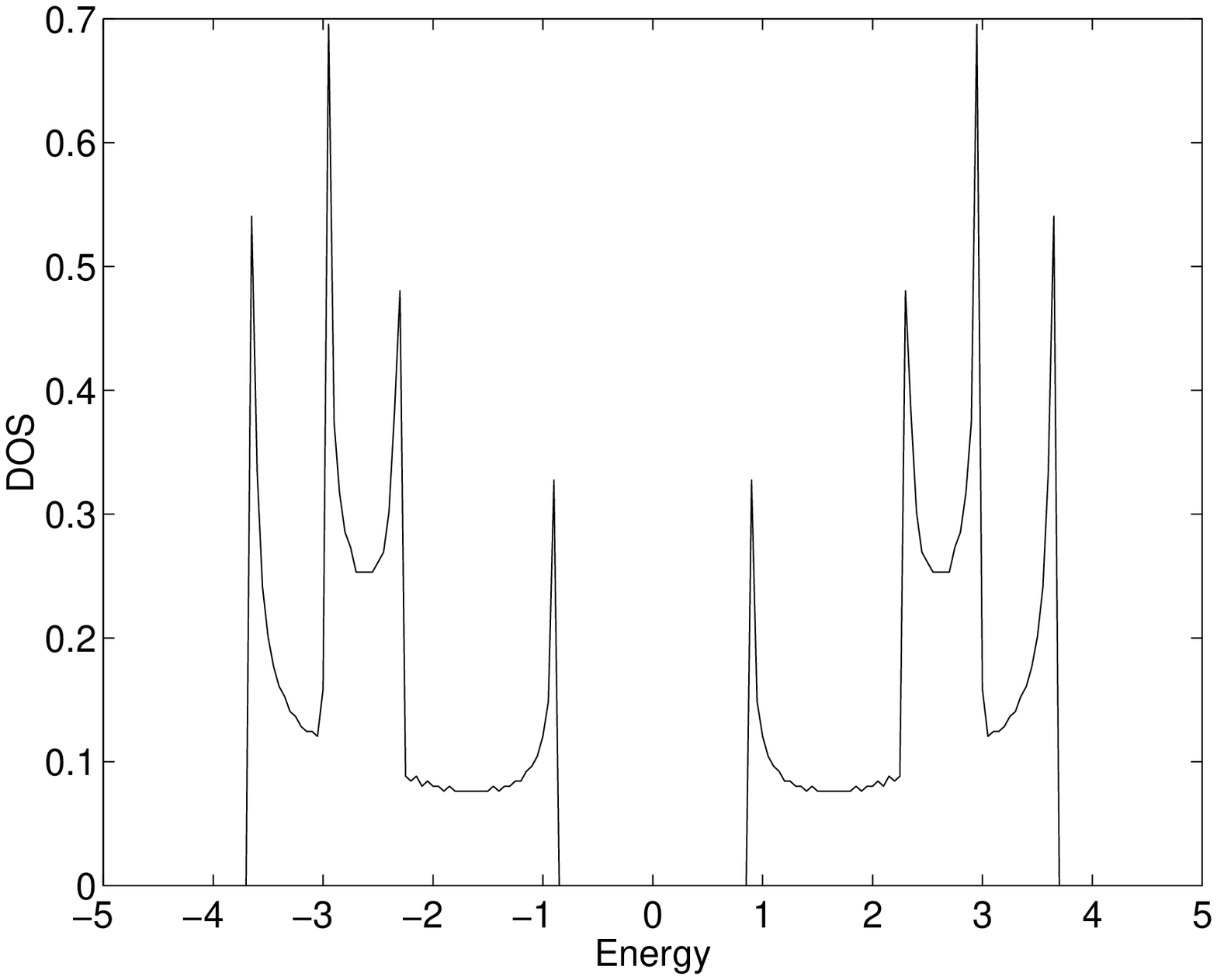}
              \end{picture}
\caption{
Density of states for 4 chains: DOS of the end chain and average over
all chains, with  $\Delta_0/t=1$ and $\mu=0$.
}
\end{figure}

Now we will apply the recursive method to evaluate the DOS of the
end chain of $N$ DWSCs. The calculation is simplified greatly
if we assume $\Delta_0 = t = 1$ (the more general case is
treated in numerical calculations).  In this special case we have
\begin{eqnarray}
H_{NN}&=&\pmatrix{
\mu+2\cos k_y +i\epsilon & -2\cos k_y\cr
-2\cos k_y & -\mu-2\cos k_y +i\epsilon \cr
}
\nonumber\\
&=&(\mu+2\cos k_y)\sigma_3-2\cos k_y\sigma_1+i\epsilon\sigma_0\nonumber\\
&\equiv& h +i\epsilon\sigma_0
\label{scham}
\end{eqnarray}
and
\[
H_{NN-1}=H_{N-1N}=\pmatrix{
1 & 1\cr
1 & -1\cr
}=\sigma_1+\sigma_3.
\]
The initial value (single chain) is
$$
\Gamma_1=(H_{11})^{-1}
={-i\epsilon\sigma_0+(\mu+2\cos k_y)\sigma_3-2\cos k_y\sigma_1
\over\epsilon^2+(\mu+2\cos k_y)^2+4\cos ^2k_y}.
$$

\begin{figure}[h]
\begin{picture}(100,200)
\leavevmode\centering\includegraphics{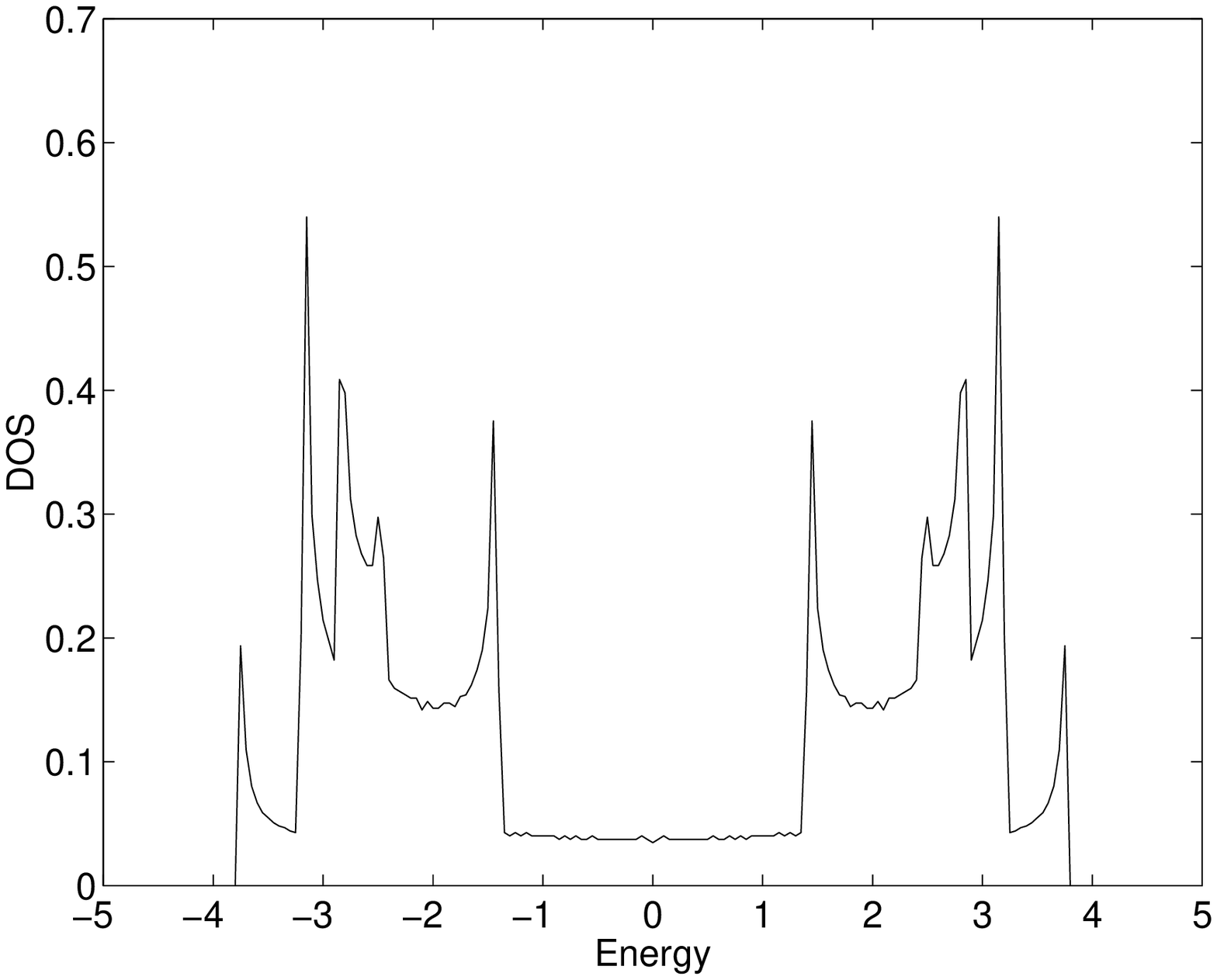}
              \end{picture}
\vskip -1cm
\begin{picture}(100,200)
\leavevmode\centering\includegraphics{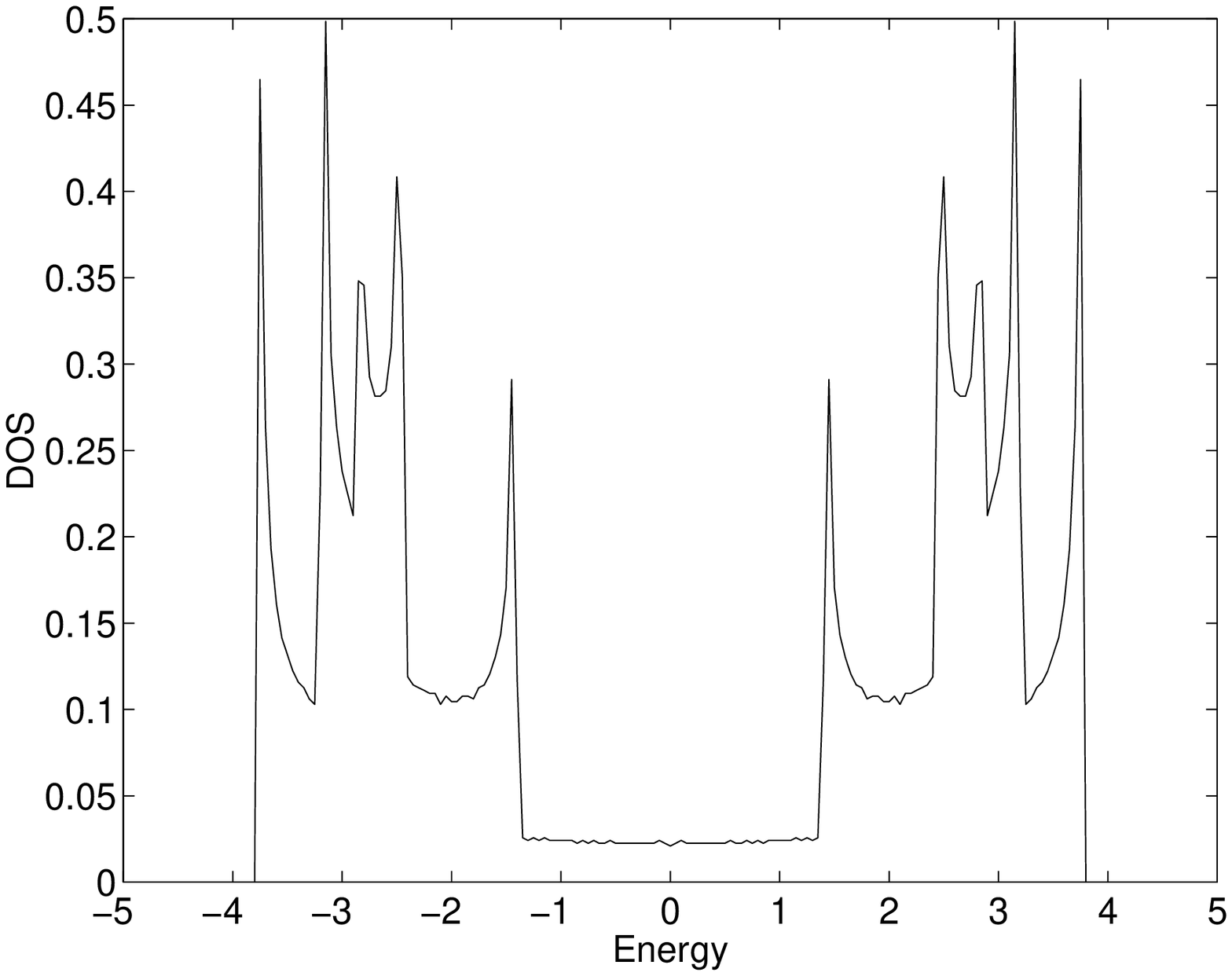}
              \end{picture}
\caption{
Density of states for 5 chains: DOS of the end chain and average over
all chains, with $\Delta_0/t=1$ and $\mu=0$.}
\end{figure}

\subsection{Few Chains}

By direct iteration of Eq. (\ref{recursion}) we can obtain the DOS
analytically for
the end chain of a system of $N$ chains, provided we restrict
consideration to $\mu=0$. Using $c=2\cos k_y$, for $N=1$ we get
$$
\rho_1(0)=\lim_{\epsilon\to0}{\epsilon\over\pi}\int_{-\pi}^\pi
{1\over\epsilon^2+2c^2}{dk_y\over2\pi}={1\over\sqrt{8}\pi}
$$
in agreement with the result in Sect. III. For $N=2$ we have
$$
\rho_2(0)=\lim_{\epsilon\to0}{\epsilon\over\pi}\int_{-\pi}^\pi
{1\over\epsilon^2+2c^2+2}{dk_y\over2\pi}.
$$
From the pole structure of the integrand,
we see that there is a gap $2E_g=2\sqrt{2}\approx 2.83$.
In the case of $N=3$ there is again a nonzero DOS at $E=0$
$$
\rho_3(0)=\lim_{\epsilon\to0}{\epsilon\over\pi}\int_{-\pi}^\pi
{\epsilon^2+2c^2+2\over(\epsilon^2+2c^2)^2+4(\epsilon^2+2c^2)}
{dk_y\over2\pi}={1\over2\sqrt{8}\pi}.
$$
and for $N=4$
$$
\rho_4(0)=\lim_{\epsilon\to0}{\epsilon\over\pi}\int_{-\pi}^\pi
{\epsilon^2+2c^2+4\over(\epsilon^2+2c^2)^2+6(\epsilon^2+2c^2)+4}
{dk_y\over2\pi},
$$
which has a gap $2E_g=2\sqrt{3-\sqrt{5}}\approx1.75$ (see Fig. 1).
In general the energy gap observed depends on the size of the
superconducting order parameter maximum $\Delta_0$ in a nonlinear
way, although relatively simple expressions may be obtained
for a small number of chains.  

The reader will notice that the even-odd effect described here differs
from the periodic boundary case, where gapless states arise for $N$
a multiple of 4.  This is expected, since the effect we are reporting
arises from self-interference of quantum wavefunctions in a nanoscale 
system.  The path-length for a constructively interfering closed path 
in the $x$-direction is $2L = 2n\pi / k_F$ for open boundary conditions,
and $L = 2n\pi / k_F$ for periodic boundary conditions.



\begin{figure}[h]
\begin{picture}(100,200)
\leavevmode\centering\includegraphics{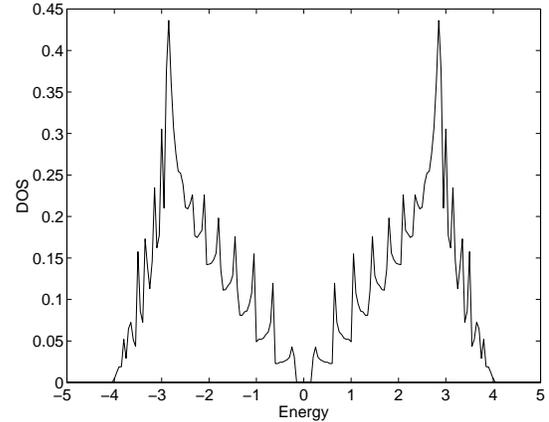}
              \end{picture}
\vskip -1cm
\begin{picture}(100,200)
\leavevmode\centering\includegraphics{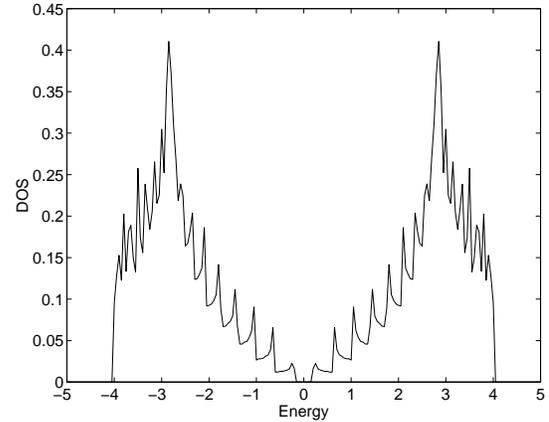}
              \end{picture}
\caption{
Density of states for 20 chains: DOS of the end chain and average over
all chains, with $\Delta_0/t=1$ and $\mu=0$.}
\end{figure}

\begin{figure}[h]
\begin{picture}(100,200)
\leavevmode\centering\includegraphics{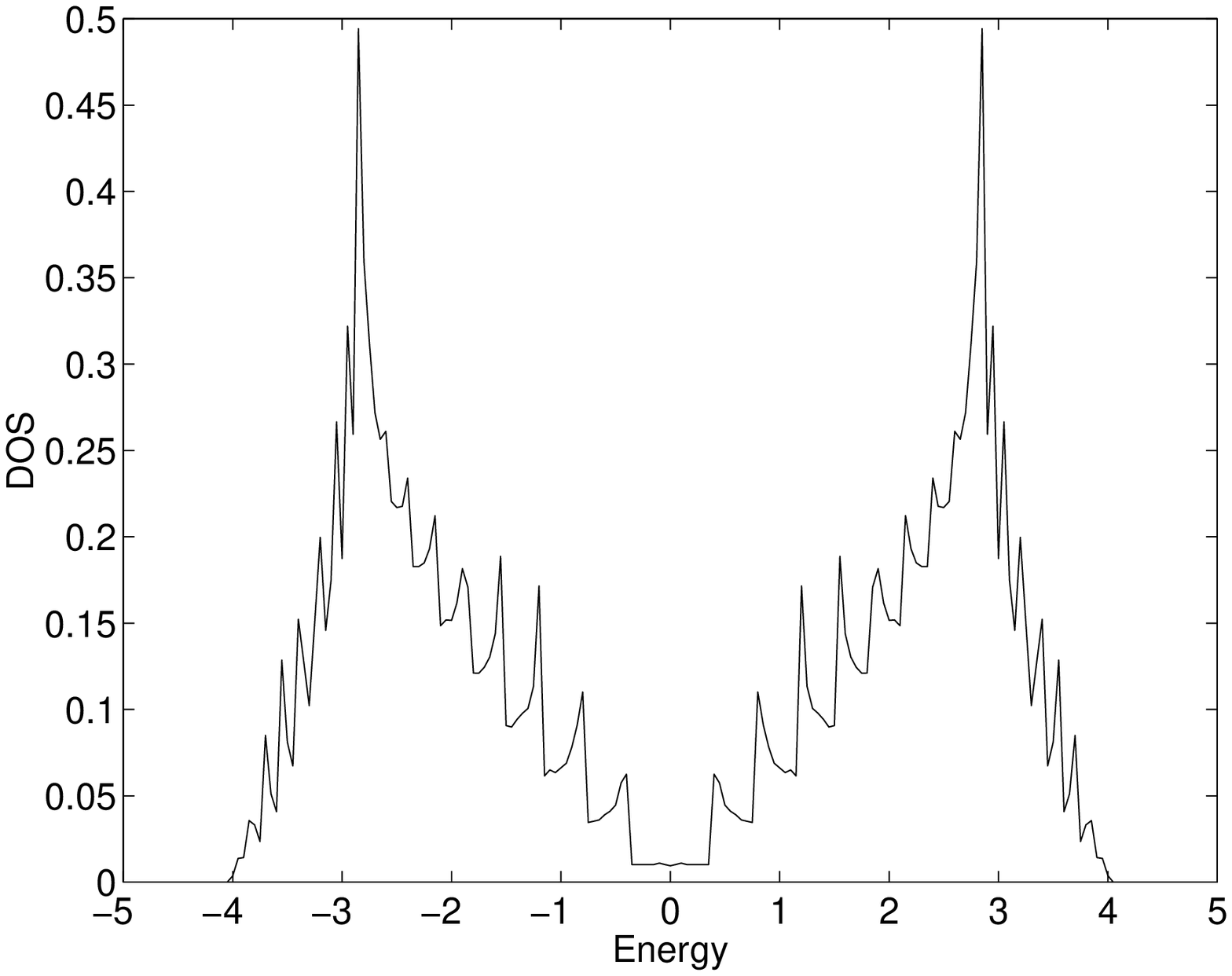}
              \end{picture}
\vskip -1cm
\begin{picture}(100,200)
\leavevmode\centering\includegraphics{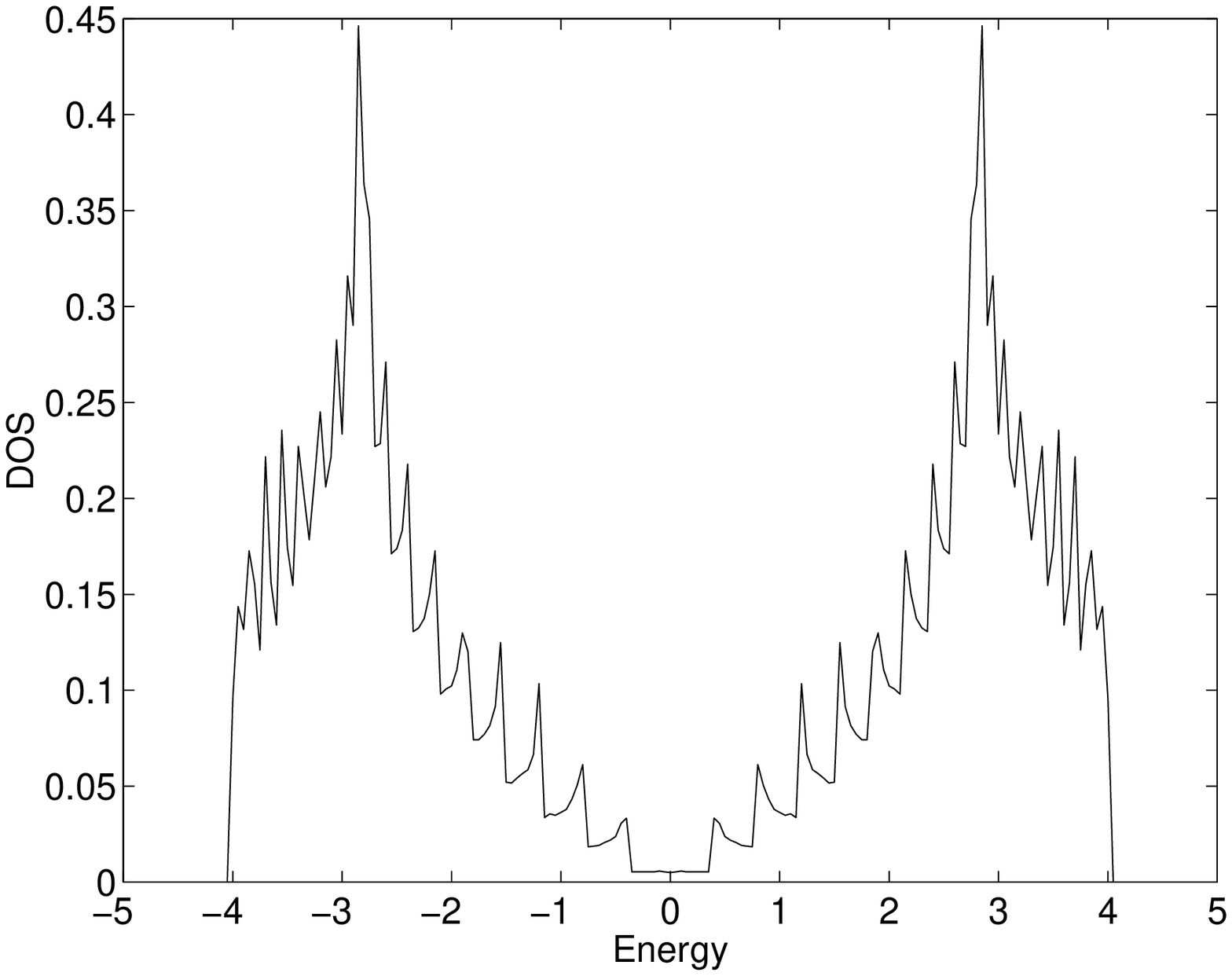}
              \end{picture}
\caption{
Density of states for 21 chains: DOS of the end chain and average over
all chains, with $\Delta_0/t=1$ and $\mu=0$.}
\end{figure}

\subsection{Many Chains: Stationary Behavior for $N\to\infty$}
While in systems mesoscopic in both directions (finite length $L$
quantum wires), even-odd parity effects in $N$ 
are known to survive the thermodynamic
limit $L\rightarrow\infty$\cite{brouwer2}, they must disappear as
$N\rightarrow \infty$ when we recover the fully $2D$ system.  
This is evident from the numerical evaluation of the DOS for
large $N$ (Figures 4 and 5), and we can see this analytically
by showing that as $N\rightarrow \infty$ $\rho(0)=0$ independent
of $N$.
\noindent

A single iteration of Eq. (\ref{recursion}) yields a relation
between the Green's functions of an even (or odd) number of chains,
respectively
\begin{eqnarray}
\Gamma_{N}&=&
(H_{NN} - H_{NN-1}
[H_{N-1N-1}\nonumber\\
&&-H_{N-1N-2}\Gamma_{N-2}H_{N-2N-1}]^{-1}
H_{N-1N})^{-1}.
\end{eqnarray}
For the DWSCs this reads
$$
\Gamma_{N}=
(H_{NN} - [{1\over2}(\sigma_1+\sigma_3)H_{N-1N-1}(\sigma_1+\sigma_3)
-\Gamma_{N-2}]^{-1})^{-1}.
$$
For constant $\mu=0$ we get from expression (\ref{scham})
$$
H_{N-1N-1}=h+i\epsilon\sigma_0
$$
and
$$
{1\over2}(\sigma_1+\sigma_3)H_{N-1N-1}(\sigma_1+\sigma_3)
=-h+i\epsilon\sigma_0
$$
such that
$$
\Gamma_{N}=
(h+i\epsilon\sigma_0 + [h-i\epsilon\sigma_0 
+\Gamma_{N-2}]^{-1})^{-1}
$$
with the initial expressions
$$
\Gamma_1={1\over\epsilon^2+2c^2}(h-i\epsilon\sigma_0)
$$
and
$$
\Gamma_2=(1+{2\over\epsilon^2+2c^2})^{-1}\Gamma_1.
$$

The symmetric $2\times2$ matrix
$h$ can be diagonalized by a orthogonal transformation, leading to diagonal
elements $\lambda_{1/2}=\pm\sqrt{2}c$.
This also diagonalizes the initial expressions $\Gamma_{1,2}$ as well as the
recurrence relation
\begin{equation}
\gamma_{N+2,j}
={1\over\lambda_j+i\epsilon+{1\over\lambda_j-i\epsilon +\gamma_{N,j}}}
\label{recursion2}
\end{equation}
with diagonal elements $\gamma_{N,1/2}$ of $\Gamma_N$. Then the recursion 
(\ref{recursion2}) has two fixed points (i.e., stationary states)
$$
{\bar\gamma}_{j}^\pm=-{z_j^*\over2}(1\pm\sqrt{1+4/|z_j|^2})
$$
with $z_j=\lambda_j+i\epsilon=-(-1)^j\sqrt{2}c+i\epsilon$. 
For $\epsilon>0$ a positive imaginary part of $\gamma_{N,j}^+$ implies
a positive imaginary part of $\gamma_{N+2,j}^+$. Since the
imaginary parts of the initial values are positive, only the fixed point
${\bar\gamma}_j^+$ can be reached in the case under consideration.

The DOS of the end chain can be calculated from the fixed point and reads
$$
\rho(0)=-{1\over\pi}\lim_{\epsilon\to0}Im\int_{-\pi}^\pi
({\bar\gamma}_1^++{\bar\gamma}_2^+){dk_y\over2\pi}
\propto\lim_{\epsilon\to0}\epsilon\ln(1/\epsilon)=0.
$$
This reflects the well-known result of the infinite 2D DWSC which
has a linear pseudogap. 
\vskip .2cm

\subsection{Realistic models}
\begin{figure}[h]
\begin{picture}(100,200)
\leavevmode\centering\includegraphics{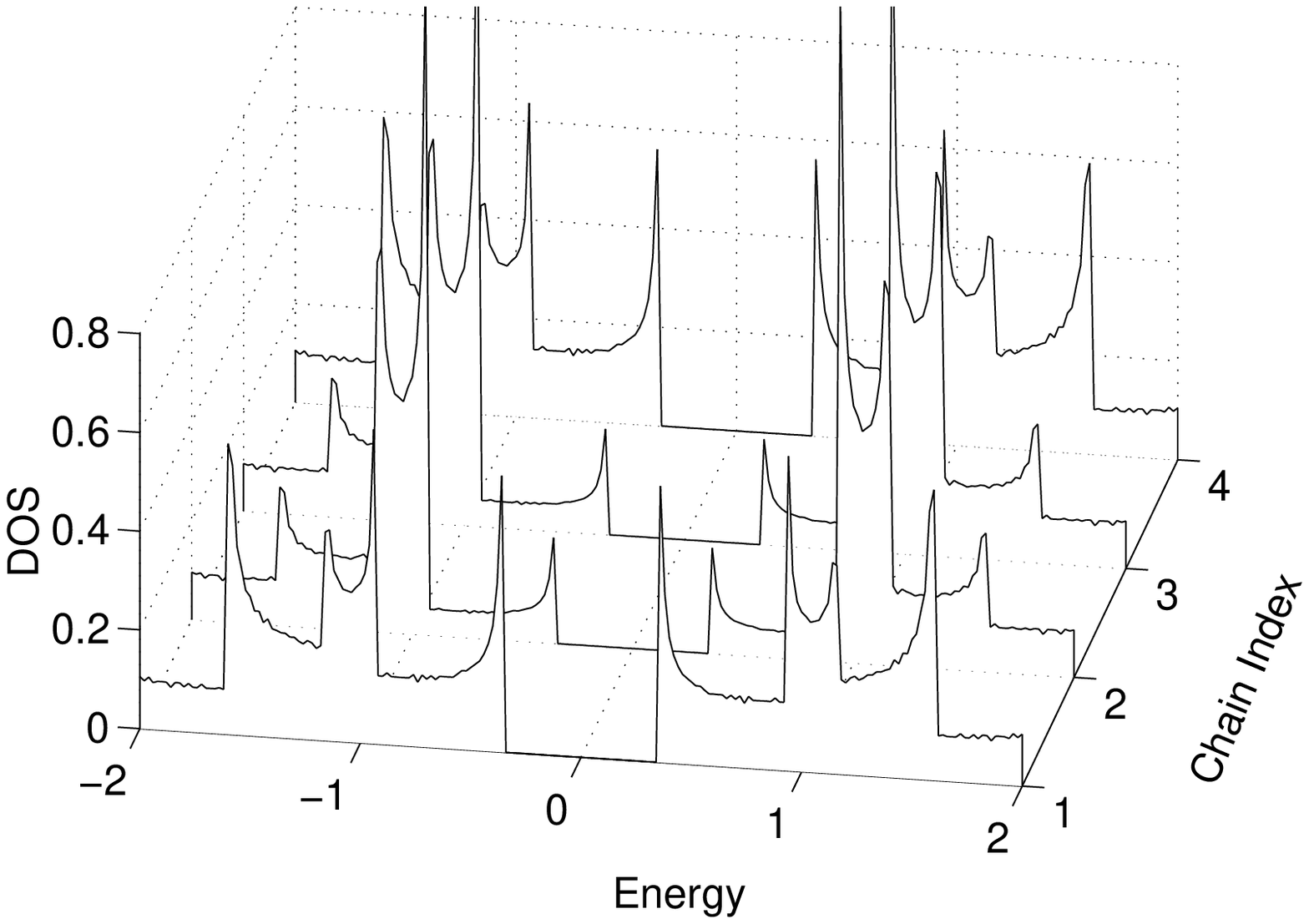}
              \end{picture}
\vskip -1cm
\begin{picture}(100,200)
\leavevmode\centering\includegraphics{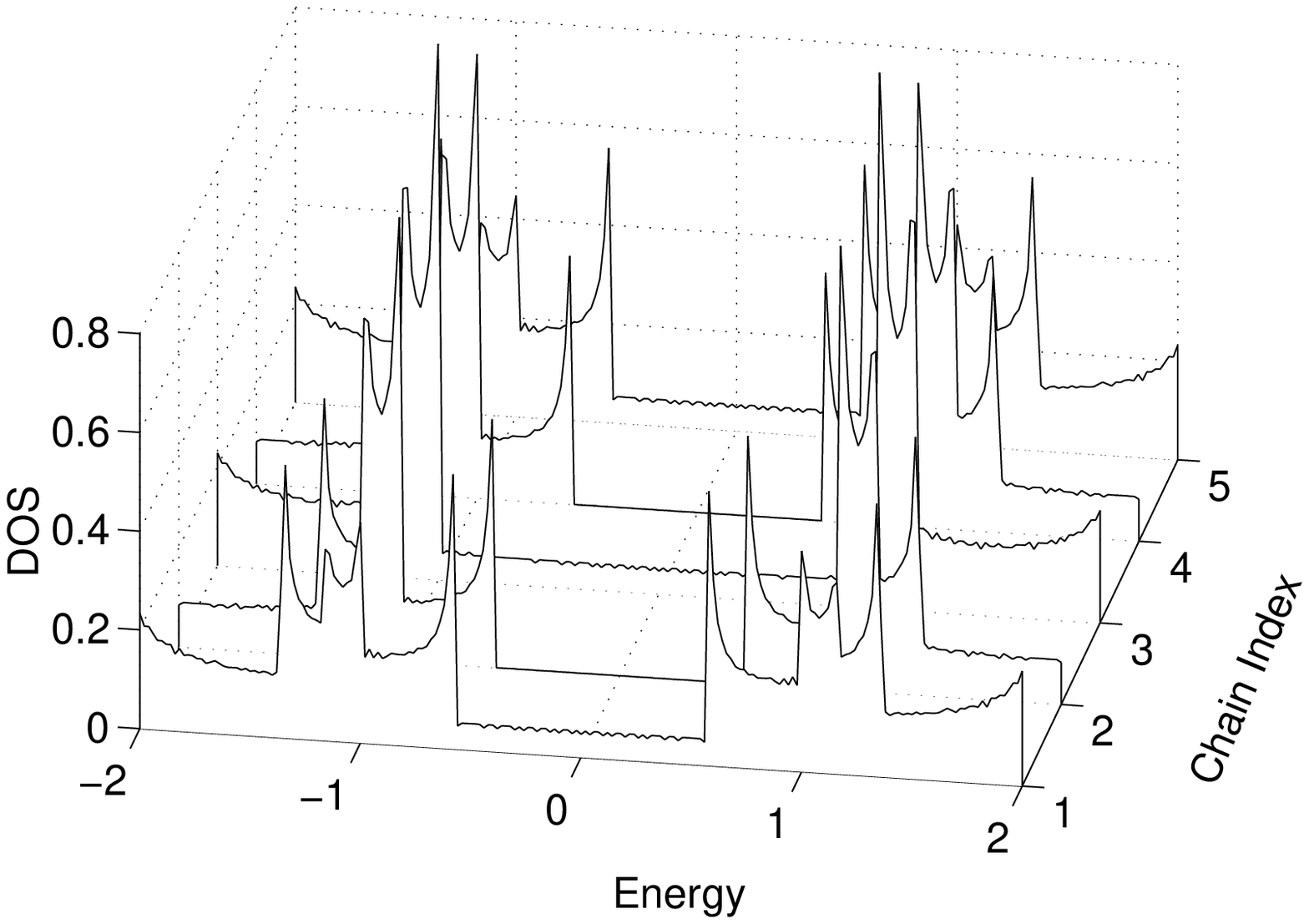}
              \end{picture}
\caption{
Density of states for four and five chains with $\Delta_0/t=0.3$ and
$\mu=0$.}
\end{figure}

The preceding discussion has assumed a half-filled tight-binding
band and $\Delta_0/t=1$; both conditions correspond to special
symmetries of the Hamiltonian  which
might be expected to influence the form of the spectrum. The assumption
$\Delta_0/t=1$ was in fact made simply in order to obtain the analytical
results discussed in sections IV A-C, and it may easily be
checked numerically (Figure 6) 
that for small values of $\Delta_0/t$ and $\mu=0$, the
qualitative even-odd effect in the chain width parity is
still obtained.   This is consistent with the qualitative results
in the case of periodic boundaries. 

More significant changes occur when $\mu\ne 0$.
In Figure 7 we see that for an
even number of chains, the full gap in all chains is preserved, while 
for an odd number, those chains which at half-filling 
had a finite density of states have now acquired a small gap.
Analytic results for this case are complicated and not
particularly enlightening, but it is clear that for
 $\mu\ll \Delta_0$, the small gap on the $x$=odd
chains in the odd $N$ case is of the order of $\mu$ itself.
This is also evident by analogy to Eq. (1), where the pole in the
denominator of the integral is shifted by the chemical potential.

\begin{figure}[h]
\begin{picture}(100,200)
\leavevmode\centering\includegraphics{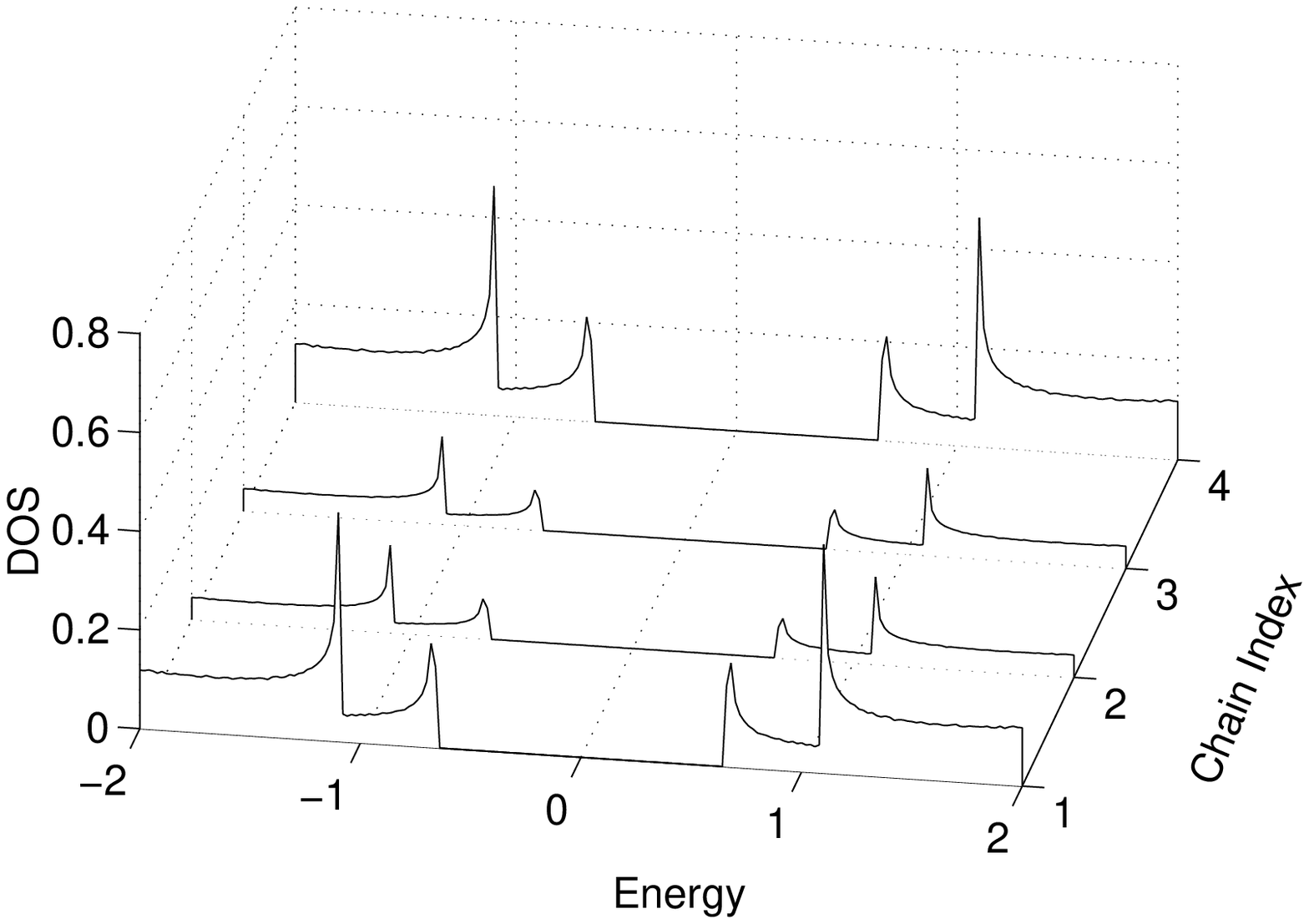}
              \end{picture}
\vskip -1cm
\begin{picture}(100,200)
\leavevmode\centering\includegraphics{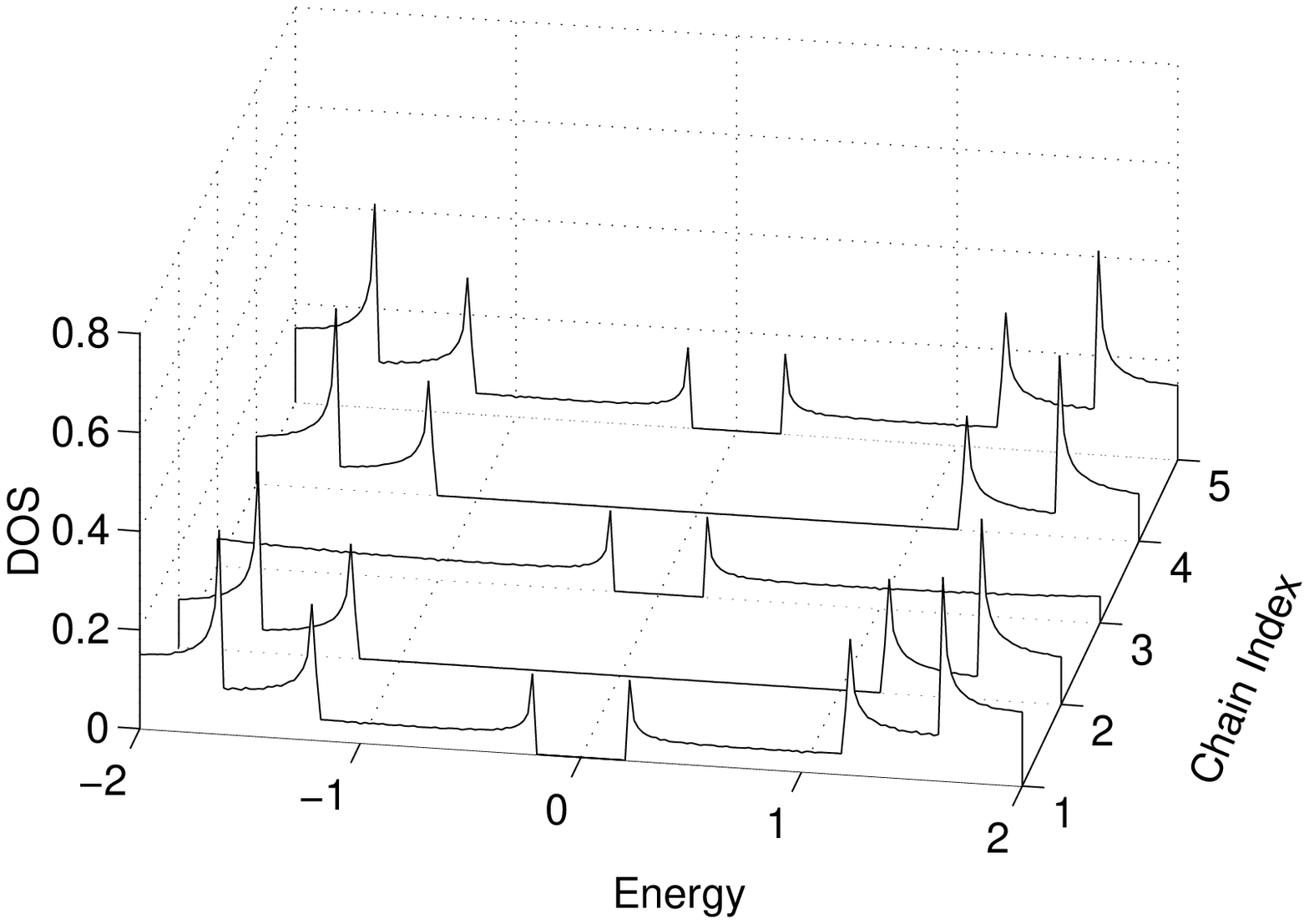}
              \end{picture}
\caption{
Density of states for four and five  chains with $\Delta_0/t=1$ and
$\mu/t=0.3$.}
\end{figure}

A more realistic parameter choice for the optimally doped cuprates would
correspond to a doping of $15\%$, or about $\mu \simeq 0.3t$ 
for the  simple tight-binding spectrum on a square lattice, and a much 
smaller gap magnitude, of order
$\Delta_0=0.1t$.  
In this case the large ``gap'' is set by $2\Delta_0$, but in 
the odd-chain systems there is a much smaller gap on alternating
chains,
in Figure 8.  Thus
the density of states  still generally exhibits a pronounced odd-even
width parity effect. 

\begin{figure}[h]
\begin{picture}(100,200)
\leavevmode\centering\includegraphics{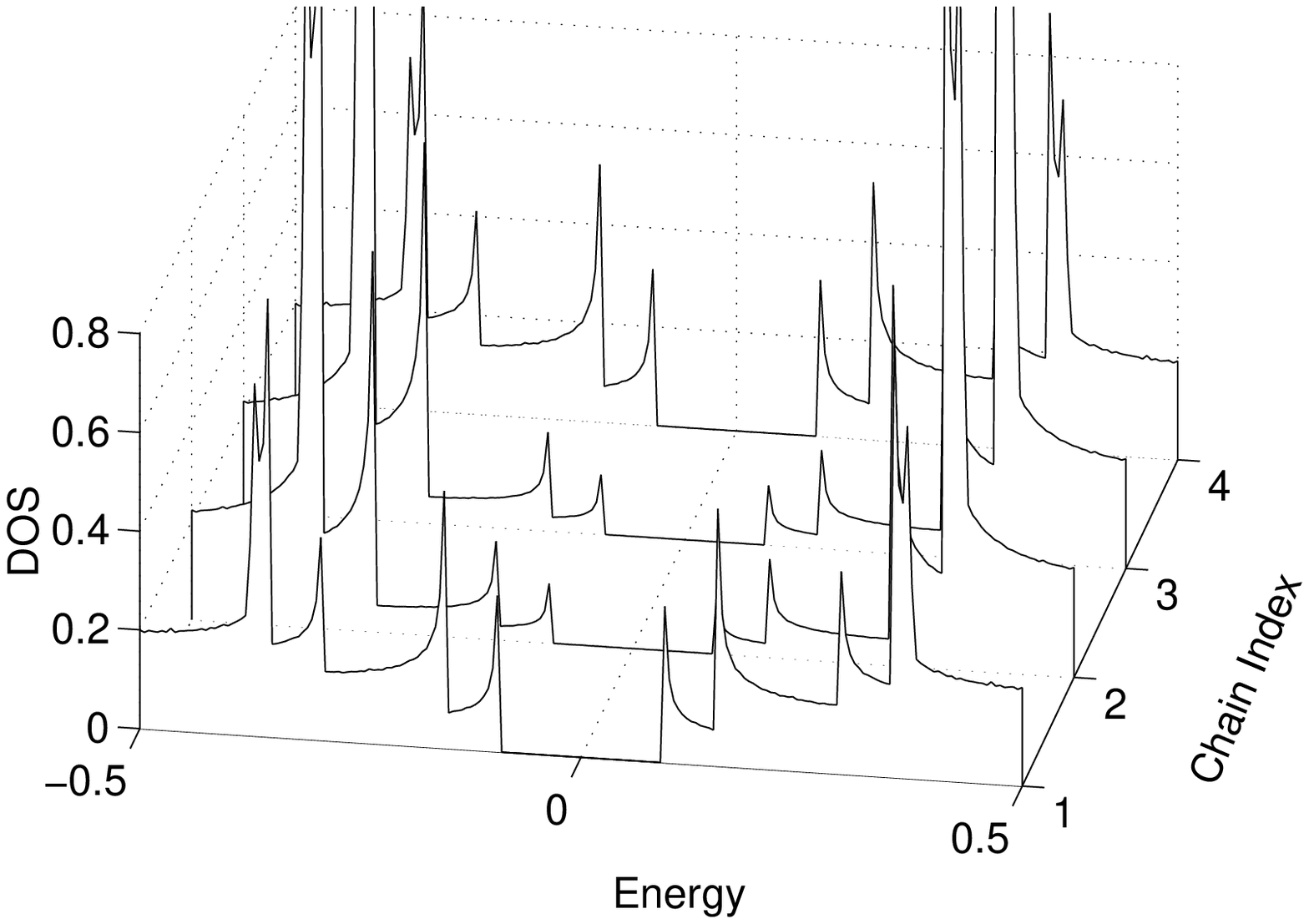}
              \end{picture}
\vskip -1cm
\begin{picture}(100,200)
\leavevmode\centering\includegraphics{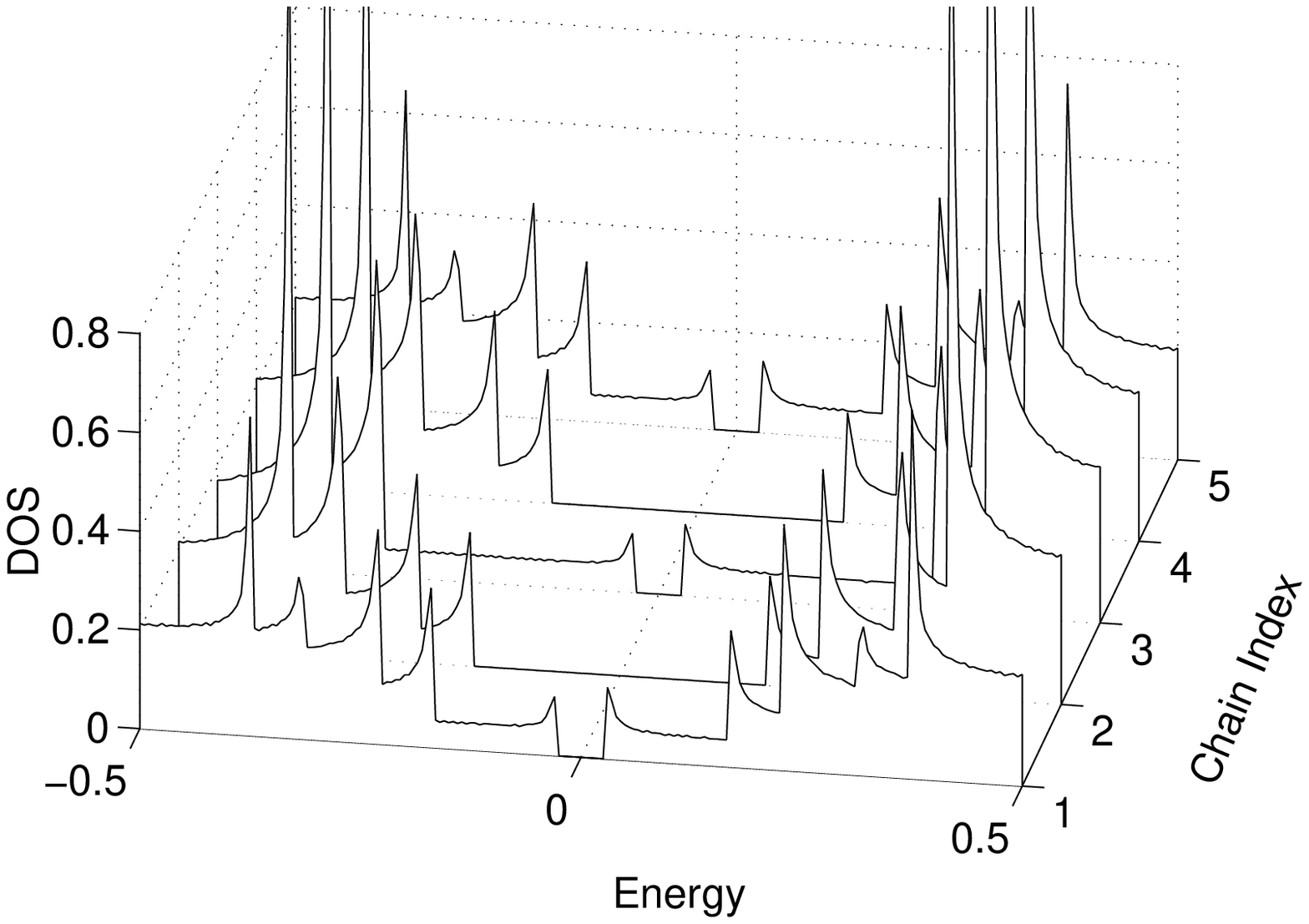}
              \end{picture}
\caption{
Density of states for four and five  chains with $\Delta_0/t=0.1$ and
$\mu/t=0.3$.}
\end{figure}

\section{Conclusions}

We have exhibited, using both analytical and numerical means,
a number parity effect in the width $N$ of a mesoscopic $d$-wave
superconducting quantum wire, wherein a finite DOS
is found at the Fermi level for odd $N$, and zero DOS (with energy gap)
for even $N$.  The result should be of some practical interest
in the not-so-distant future for nanoscale $d$-wave wire structures.
Superconductivity in single layers of the cuprates has already been
demonstrated, and it seems plausible  that a sample of controlled
width might be fabricated, and  that an STM experiment on
such a sample would be able to observe the effects we predict.  

It is interesting to note some unusual features of the system we
treat here, to our knowledge for the first time, in 
the presence of impurities.  First of all,
although the superconducting order parameter has $d$-wave symmetry,
the system may (for $N$ even) have a full gap in the excitation spectrum.
In such a situation, one may ask what is the effect of isolated
impurities added to the system, and at least naively would obtain
single-impurity $d$-wave like bound states in  this gap without
broadening arising from coupling to the quasiparticle continuum.
Since the density of states in the odd-$N$ systems has 
an oscillatory behavior across the sample width, the existence
and lifetime of these states is expected to depend sensitively
on their location in the wire as well.

In addition, the observation we make here may be of interest 
to the study of the influence of disorder on the DOS of 
 fully 2D $d$-wave superconductors,
a subject which has received intense attention recently.\cite{2Ddos}
The result appears to be that the DOS is generically zero in a $2D$ $d$-wave
superconductor  at zero energy, but can be constant or divergent
in  cases manifesting special symmetries. At the same time,
  Brouwer et al.\cite{brouwer2}
have shown in disordered quasi1D systems with chiral symmetry that $\rho(0)$
is zero or divergent according to the number parity of the wire width.
It appears  that the states in odd or even chain systems
are modified in different ways by the combined effects of level
repulsion and symmetry.  Understanding the differences in the
odd or even approach to the thermodynamic limit in the presence
of disorder may give insight into the physics of localization and
DOS supression in the 2D $d$-wave case, which is still poorly understood.
We will address these questions in a subsequent work.

{\it Acknowledgements}

This work is supported by NSF grants DMR-9974396 and
INT-9815833, and an exchange grant from DAAD. 
  The authors are grateful to J. von Delft and 
C. Mudry for useful discussions.  PH is grateful to the Institute
for Theoretical Physics for its hospitality during work on the manuscript.


\begin{references}
\bibitem{Ralph1} D.C. Ralph, C.T. Black, and M. Tinkham, PRL 74
3241
(1995).
\bibitem{vonDelft} F. Braun and J. von Delft, Phys. Rev. Lett. 81, 4712 (1998).
\bibitem{Zaikin} A.D. Zaikin, D.S. Gobulev, A. van Otterlo,
and G.T. Zimanyi,  Phys. Rev.
   Lett. 78, 1552 (1997).
\bibitem{2Ddos} A. A. Nersesyan, A. M. Tsvelik, and  F. Wenger,
Nucl. Phys. B {\bf 438}, 561 (1995); Phys. Rev. Lett. {\bf 72}, 2628 (1994);
 K. Ziegler, M.H.  Hettler, and P.J. Hirschfeld,
Phys. Rev. Lett.  77, 3013 (1996); Phys. Rev. B 57, 10825 (1998);
T. Senthil, Matthew P. A. Fisher, Leon Balents,
and Chetan Nayak, Phys. Rev. Lett. {\bf 81} 4704 (1998); T. Senthil
and Matthew P. A. Fisher, Phys. Rev. B60, 6893  (1999);
 C. P\'epin and P.A. Lee, Phys. Rev. B {\bf 63} 54502 (2001); 
W.A. Atkinson, P.J. Hirschfeld, A. H. MacDonald, and K. Ziegler,
Phys. Rev. Lett. {\bf 85}, 3926 (2000).
\bibitem{nanotube} A. Kleiner and S. Eggert, cond-mat/0007244.


\bibitem{haldane}
F.D.M. Haldane, Phys. Rev. Lett. 61, 1029 (1988).
\bibitem{fradkin}
E. Fradkin, Field Theory of Condensed Matter Systems,
Addison-Wesley, Redwood City (1991).


\bibitem{miller}
J. Miller and J. Wang, Phys. Rev. Lett. {\bf 76}, 1461 (1996)

\bibitem{brouwer1}
P.W. Brouwer, C. Mudry, B.D. Simons, and A. Altland, Phys. Rev. Lett.
{\bf 81}, 862 (1998)

\bibitem{brouwer2}
P.W. Brouwer, C. Mudry, and A. Furusaki,
Phys. Rev. Lett. 84, 2913 (2000)

\bibitem{sauls} M. Fogelstrom, D. Rainer, and J. A. Sauls, Phys.
Rev. Lett. {\bf 79}, 281 (1997).

\end{references}
\end{document}